\documentclass[a4paper,twocolumn,11pt]{quantumarticle}
\pdfoutput=1
\usepackage[utf8]{inputenc}
\usepackage[english]{babel}
\usepackage[T1]{fontenc}
\usepackage{amsmath}
\usepackage{hyperref}
\usepackage[numbers,sort&compress]{natbib}
\usepackage{tikz}
\usepackage{braket}
\usepackage{lipsum}
\usepackage{qcircuit}
\usepackage{amssymb}

\newtheorem{lemma}{Lemma}
\newtheorem{corollary}{Corollary}

\usepackage[utf8]{inputenc}
\usepackage[noend,noline,ruled,linesnumbered]{algorithm2e}
\usepackage{soul}
\usepackage{placeins}

\usepackage[utf8]{inputenc}

\newcommand{\Rbb}{\mathbb{R}}

\newcommand{\Acal}{\mathcal{A}}
\newcommand{\Bcal}{\mathcal{B}}
\newcommand{\Ccal}{\mathcal{C}}

\newcommand{\Lcal}{\mathcal{L}}

\begin{document}

\title{Quantum circuit synthesis with qudit phase gadget method}

\author{Shuai Yang}
\address{Hangzhoudianzi University, Hangzhou, China}
\author{Lihao Xu}
\author{Guojing Tian}
\address{Institute of Computing Technology, Chinese Academy of Sciences, 100190 Beijing, China}
\address{University of Chinese Academy of Sciences, 100049 Beijing, China}
\author{Xiaoming Sun}
\address{Institute of Computing Technology, Chinese Academy of Sciences, 100190 Beijing, China}
\address{University of Chinese Academy of Sciences, 100049 Beijing, China}
\email{sunxiaoming@ict.ac.cn}
\maketitle

\begin{abstract}
   Current quantum devices have unutilized high-level quantum resources.
  More and more attention has been paid to the qudit quantum systems with larger than two dimensions to maximize the potential computing power of quantum computation. Then, a natural problem arises: How do we implement quantum algorithms on qudit quantum systems? 
In this work, we propose a novel qudit phase gadget method for synthesizing the qudit diagonal unitary matrices.
This method is suitable for the Noisy Intermediate-Scale Quantum (NISQ) and fault-tolerant eras due to its versatility in different connectivity architectures and the optimality of its resource consumption. The method can work on any connectivity architecture with asymptotic optimal circuit depth and size. For a 10-qutrit diagonal unitary, our algorithm reduces the circuit depth form $\sim 1\times 10^5$ to 500 with 300 ancillary qutrits.
Further, this method can be promoted to different quantum circuit synthesis problems, such as quantum state preparation problems, general unitary synthesis problems, etc.
\end{abstract}

\section{Introduction}
Quantum computation has made significant progress in the past few decades~\cite{Shor1994Polynominal,grover1996fast}, since Feynman first proposed the concept of quantum computation~\cite{feynman1982simulating}. To implement quantum algorithms on quantum devices, an essential task is to decompose quantum algorithms into elementary quantum gates with as few resources (including classical and quantum resources) as possible. The process described above is typically called quantum circuit synthesis~\cite{barenco1995elementary,di2016parallelizing}.


Research on quantum circuit synthesis algorithms has a long history. In 1995, Barenco et al. first proposed the quantum circuit synthesis problem and introduced a synthesis algorithm for $n$-qubit unitary matrices, which takes $O(n^3 4^n)$ elementary gates~\cite{barenco1995elementary}. This result was later improved to $O(4^n)$ by Bergholm et al. in 2005~\cite{bergholm2005quantum}. Then in 2006, Shende et al. designed a novel synthesis algorithm based on the Schmidt decomposition. For a general $n$-qubit unitary, it only needs $\frac{23}{48}4^n$ CNOT gates and $O(4^n)$ elementary single-qubit gates~\cite{shende2006synthesis} to complete the synthesis. Later, in 2023, Sun et al. focused on the circuit depth. They proposed the qubit phase gadget method to reduce the circuit depth to $O(4^n/(n+m)+n2^n)$ with $m$ ancillary qubits~\cite{sun2021asymptotically}.

For some specific unitary families, specialized algorithms have also been designed. One of the most famous examples is the quantum state preparation problem. The input of $n$-qubit ($n$-qudit) quantum state preparation problem is a $2^n$ ($d^n$) complex vector $v=(v_0,v_1,\cdots)$. And the target is to find the gate sequence $L$ such that $L\ket{0}=\sum_k v_k\ket{k}$.
For $n$-qubit quantum state preparation, Bergholm et al. came up with a synthesis algorithm that takes $\frac{23}{24}2^n$ CNOT gates and $O(2^n)$ elementary single-qubit gates~\cite{shende2006synthesis}. In 2023, Sun et al. first reduced the circuit depth to poly$(n)$~\cite{sun2021asymptotically}. They also gave an analysis for the lower bound of these two problems, and their algorithm is not only asymptotically optimal in circuit size but also in circuit depth.

In recent years, quantum computation has been in the Noisy Intermediate-Scale Quantum (NISQ) era ~\cite{arute2019quantum,preskill2018quantum}. In the NISQ era, quantum resources, such as the fidelity and decoherence time of quantum gates, are limited. The core task of the quantum synthesis problem in NISQ is to fully use quantum resources to implement as large-scale a problem as possible. A frequently overlooked quantum resource is the high-level state.
Most of the existing research on quantum computing is focused on two-level quantum systems; the higher dimensions are widely presented in existing physical systems, such as photonic systems~\cite{Chi2022}, ion traps~\cite{PhysRevA.67.062313,Ringbauer2022}, and superconducting devices~\cite{Blok2021,Yurtalan2020}. These temporarily idle energy levels may reduce the quantum costs of quantum computation. For example, to encode an $N$ items array into the quantum device, it will need $\lceil\log N\rceil$ qubits in a qubit system, but in a qudit system, the number is only $\lceil\log_d N\rceil$.

Compared to the rich quantum circuit synthesis of two-level systems, there is relatively little research on quantum circuits for multi-level systems. Current research on quantum circuit synthesis for multi-level quantum systems focuses on several aspects, including quantum state preparation~\cite{Bullock2004AsymptoticallyOQ}, general unitary synthesis~\cite{Bullock2004AsymptoticallyOQ,zi2023optimal,Ivanov2006Engineering,Prakash2018Normal}, and reversible Boolean function synthesis~\cite{zi2023optimal,Yeh2022constru}. These works focus more on reducing the quantum size than optimizing the quantum circuit depth. For quantum state preparation and general unitary synthesis, the gate count is about $O(d^n)$,$O(d^{2n})$, corresponding.

It is difficult to reduce the depth of the qudit circuit, since there is no parallel framework for it. Due to the diversity of operations in the qudit system, the framework of the qubit system can not be used directly in the qudit system. 
To further reduce the circuit depth without increasing the circuit size, we propose a new method, called the qudit phase gadget method, for the qudit circuit synthesis problem in this work. Similarly to the qubit phase gadget method, we prove that there is a transformation between the diagonal unitary and qudit phase gadget circuits. To prove this fact, we first construct the transform between the diagonal unitary and qudit phase gadget circuit parameters. Then, we construct a universal qudit phase gadget circuit that contains all the phase gadgets. This construction reduces the number of qudit gates to $O(d^{n-1} \log{d})$. In addition, our method can significantly improve the construction of various synthesis problems. 

\textbf{Qudit state preparation---} Bullock et al. first gave a club-sequence-based method to prepare $n$-qudit state preparation~\cite{Bullock2004AsymptoticallyOQ}. Their synthesis algorithm can reduce the number of two-qudit gates to $(d^n-1)/(d-1)$ with several ancillary qudits. Their algorithms do not care about the circuit depth. The circuit depth of their method is also $O(d^{n-1})$. They also present an analysis of the lower bound of qudit state preparation. With Sard's theorem, almost all $n$-qudit states can not be prepared by a $o(d^{n-4})$-depth circuit. 

Our method provides a trade-off between the number of ancillary qudit and the circuit depth. We construct an $O(d^{n-1}\log d)$-size and $O(\frac{d^{n-1}\log d}{(n+m)}+n^2\log d)$-depth quantum circuit for $n$-qudit {(unitary)} and $m$ ancillary qudits. We also prove the depth lower bound of the qudit state preparation problem. Our algorithm is asymptotically optimal for a large range of ancillary qudits.

\textbf{Qudit general unitary synthesis---}Also, Bullock et al. gave a Householder reflection based on the method to synthesize arbitrary $n$-qudit unitary. The circuit size and circuit depth is both $O(d^{2n+4})$ in their analysis~\footnote{In their analysis, the poly$(d)$ term can be ignore.} and takes $\lceil (n-2) /(d-2)\rceil$ ancillary qudits. They also showed an $\Omega(d^{2n-4})$ lower bound for the number of two-qudit gates of the general unitary synthesis problem. Zi et al. further reduced the number of ancillary qudit to 1 in very recent years\cite{zi2023optimal}. 

Our method not only reduces the circuit size to $O({d^{2n-1}\log d})$ but gives an (almost) asymptotically optimal trade-off between the circuit depth and the number of ancillary qudits. The circuit depth of general unitary synthesis problem is $O(\frac{d^{2n-1}\log d}{n+m}+n^2d^n\log {d})$ with $m$ ancillary qudits. The lower bounds of qudit state preparation and general unitary synthesis are obtained by counting and light cone methods.

\textbf{Organization} The remaining part of this paper will be organized as follows. In section \ref{sec:preliminaries}, we will introduce the notation used in this paper and briefly review the qubit phase gadget method. Then, in section \ref{sec:quditphasegadget}, we define the qudit phase gadget by analogy with the definition of qubit phase gadget and provide an extensive discussion of the properties of the qudit phase gadget, as well as how to use it to synthesize diagonal unitary, later in section \ref{sec:application}, we demonstrate how to use the qudit phase gadget method to solve quantum state preparation problem and qudit general unitary synthesis problem. The circuit depth is significantly reduced by applying our method. Then, we conclude in section \ref{sec:conclusion}.

\section{PRELIMINARIES}
\label{sec:preliminaries}
\subsection{NOTATION}
\label{sec:notation}
This section will introduce some basic notations used in this paper.

Let $[x]$ denote the set $\{0,1,2,3,\cdots,x-1\}$. And let $\langle a,b\rangle$ be the inner product of $a,b$. 
We first define the $[d]$-string $s\in[d]^*$ as the string consists of number in $[d]$. Then let the inner product of two $[d]$-strings $s_1,s_2$ be $\langle s_1,s_2\rangle=\sum_i s_{1i}s_{2i}$.

\begin{figure}[!htbp]
    \begin{center}
    \mbox{
        \Qcircuit @C=1.2em @R=1.4em {
        \lstick{\ket{q_{l_1}}} & \ctrl{6} & \qw & \qw & \qw \\
        \lstick{\vdots~~~}& \qw & \qw & \qw & \qw  \\
        \lstick{\ket{q_{l_2}}} & \qw & \ctrl{4} & \qw & \qw \\
        \lstick{\vdots~~~} & \qw & \qw & \qw & \qw  \\
        \lstick{\ket{q_{l_{|l|-1}}}} & \qw& \qw & \ctrl{2} & \qw\\
        \lstick{\vdots~~~} & \qw &\dstick{~~~~~~\cdots} \qw&  \qw&\qw\\
        \lstick{\ket{q_{l_{|l|}}}}&  \targ& \targ & \targ & \qw\\
        }}
        
    \end{center}
    \caption{\centering{The linear CNOT circuit for a qubit sequence $l=\{l_1,l_2,\cdots,l_{\vert l\vert}\}$.}}
    \label{fig:linearCNOTcircuit}
\end{figure}
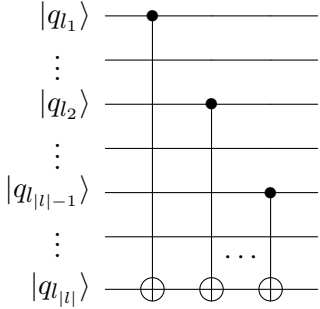

For the qubit quantum system, $CNOT_{a,b}$ denotes the CNOT gate whose control qubit is $a$ and the target qubit is $b$, and the function of CNOT gate is  $CNOT_{a,b}\ket{q_a}\ket{q_b}=\ket{q_a}\ket{q_a\oplus q_b}$, {where $\ket{q_a}, \ket{q_b}$ is the state of qubit $a,b$ correspondingly}. A linear CNOT circuit $\Lcal_l$ of a qubit sequence $l=\{l_1l_2,\cdots,l_{|l|}\}$ {is defined as} the CNOT circuit with $\vert l\vert-1$ CNOT gates as shown in Fig.\ref{fig:linearCNOTcircuit}. These CNOT gates share the same target qubit $l_{\vert l\vert}$. The control qubits are the sequence $\{l_1,l_2,\cdots,l_{|l|-1}\}$. The function of the linear CNOT circuit generates $f(q)=q_{l_1}\oplus q_{l_2}\oplus\cdots\oplus q_{l_{|l|}}$ on the target qubit, that is $\ket{c}\to\ket{f(q)\oplus c}$. This circuit can regard as a ``fan-in'' circuit.

For the $d$-dimension quantum system with $d>2$, in this paper, we will use the $Z_d, X_d, R_{d,z}(\theta)$ and $SUM_d$ gates of the qudit system. Similar to the $R_z(\theta)$ on the qubit system, the $R_{d,z}(\theta)=\mbox{diag}(1,e^{i\theta_{1}},e^{i\theta_{2}},\cdots,e^{i\theta_{d-1}})$, where $\theta=(\theta_1,\theta_2,\cdots,\theta_{d-1}).$ The $Z_d$ gate of qudit system is $\mbox{diag}(1,\omega,\cdots,\omega^{d-1}),\omega=e^{2\pi i/d}$. On the qubit system, the $X_d$ gate and the $Z_d$ gate {will be degenerated to bit flip operator $X$ and phase flip operator $Z$.}  $SUM_{d,a,b}$ denotes the gate whose control qudit is $a$ and the target qudit is $b$, and $SUM_{d,a,b}\ket{q_a}\ket{q_b}=\ket{q_a}\ket{q_a \oplus_d q_b}$, {where the $\oplus_d$ means modular $d$ addition}. For a string $s\in [d]^n$, $l=\{i|s_i\ne 0\}$  and let the item in $l$ be ordered from small to large. Then $\mathcal{L}_s$ denotes {the circuit $\Pi_{i=1}^{|l|-1} SUM^{s_{l_i}}_{d,l_i,l_{|l|}}$ including $SUM_d$ gates.}
The function of $\mathcal{L}_s$ generates $f(q)=\langle{s,q}\rangle$ on the target qubit.
In this paper, different permutations will lead to different $\mathcal{L}_s$; it will not influence the correction of the circuit since the change on the qubits will be restored later. When the $s$ contains only one non-zero item, the $\mathcal{L}_s=I$.

\begin{figure}[!htbp]
    \begin{center}
    \mbox{
        \Qcircuit @C=1.2em @R=1.4em {
        \lstick{\ket{q_{l_1}}} & \gate{s_{l_1}}\qwx[6] & \qw & \qw & \qw\\
        \lstick{\vdots~~~} & \qw & \qw & \qw & \qw \\
        \lstick{\ket{q_{l_2}}} & \qw & \gate{s_{l_2}} \qwx[4] & \qw & \qw \\
        \lstick{\vdots~~~} & \qw & \qw & \qw & \qw\\
        \lstick{\ket{q_{l_{\vert l\vert-1}}}} & \qw & \qw & \gate{s_{l_{|l|-1}}}\qwx[2] & \qw \\
        \lstick{\vdots~~~} & \qw &\qw &\dstick{\cdots~~~~~~} \qw & \qw\\
        \lstick{\ket{q_{l_{\vert l\vert}}}} &  \targ & \targ & \targ & \qw\\
        }}
    \end{center}
    \caption{{The qudit linear $SUM_d$ circuit $\mathcal{L_s}$. The number in the square means we repeat the $SUM_d$ circuit $s_{l_i}$ times.}}
    \label{fig:quditLs}
\end{figure}
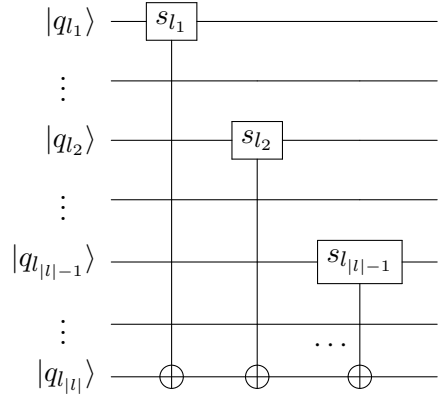

\subsection{Qubit phase gadget method}
\label{sec:qubitphasegadget}
Phase gadget is a helpful gadget widely used in circuit synthesis and quantum chemistry. The function of the phase gadget is to add phase on some specific computational basis {state}. This section will systematically introduce the phase gadget on the qubit quantum system.

On an $n$-qubit system, we use the notation $P_{\alpha,s}$ to denote the phase gadget with two parameters $\alpha \in \Rbb, s\in\{0,1\}^n$.
A phase gadget $P_{\alpha,s}$ is a circuit which can transfer the {$n$-qubit} state $\ket{x}= \ket{x_1\cdots x_n}$ to $e^{i\alpha\langle x,s\rangle}\ket{x}$ for any $x\in [2^{n}]$, {where the element $s_t=1$ iff $t\in\{j_1,j_2,\cdots,j_\ell\}$.}
It completes the transform from $\ket x = \ket{x_1\cdots x_k}$ to $e^{i \langle s,x\rangle \alpha}\ket{x}$, 
Thus the $P_{\alpha,s}$ can also be rewritten as $P_{\alpha,s}=e^{i\alpha \Pi_{s_j=1} Z_{j}}$, which is widely used in quantum Hamiltonian simulation. 
The researchers have well-studied the phase gadget, and there are several ways to synthesize it. The most common method is divide the phase gadget $P_{\alpha,s}$ into 3 parts, i.e., $C_1,R_z(\alpha/2),C_2$. The circuit $C_1,C_2$ are two CNOT circuit and $C_1=\mathcal{L}_s$ and $C_2=C_1^{-1}$. The following Fig.\ref{fig:cnotcircuit} describes the method to synthesis phase gadget.

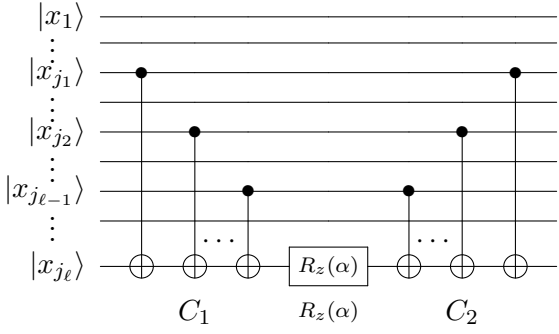
\begin{figure}[!htbp]
    \begin{center}
    \mbox{
        \Qcircuit @C=1em @R=0.9em {
        \lstick{\ket{x_1}} & \qw & \qw & \qw & \qw & \qw & \qw & \qw& \qw\\
        \lstick{\vdots~~~} & \qw  & \qw & \qw & \qw & \qw & \qw & \qw & \qw\\
        \lstick{\ket{x_{j_1}}} & \ctrl{6} & \qw & \qw & \qw & \qw & \qw & \ctrl{6}& \qw\\
        \lstick{\vdots~~~} & \qw & \qw & \qw & \qw & \qw & \qw & \qw& \qw\\
        \lstick{\ket{x_{j_2}}} & \qw & \ctrl{4} & \qw & \qw & \qw & \ctrl{4} & \qw& \qw\\
        \lstick{\vdots~~~} & \qw & \qw & \qw & \qw & \qw & \qw & \qw & \qw\\
        \lstick{\ket{x_{j_{\ell-1}}}} & \qw & \qw & \ctrl{2} & \qw & \ctrl{2} & \qw & \qw& \qw\\
        \lstick{\vdots~~~} & \qw &\qw &\dstick{\cdots~~~~~~} \qw & \qw & \qw &\dstick{\cdots~~~~~~} \qw & \qw & \qw\\
        \lstick{\ket{x_{j_\ell}}} &  \targ & \targ & \targ & \gate{ \scriptstyle R_z(\alpha)}& \targ &\targ & \targ & \qw\\
        &  & C_1 &  & { \scriptstyle R_z(\alpha)}&  &C_2 &  & \\
        }}
    \end{center}
    \caption{\centering{The phase gadget.}}
    \label{fig:cnotcircuit}
\end{figure}

Since each phase gadget can be decomposed into three parts, where the first and third parts are CNOT circuits, we can reduce the quantum resource when we apply more than one phase gadget. Assume there is two phase gadgets $P=C_1 R_z(\alpha/2)C_2$ and $P'=C'_1 R_z(\alpha'/2)C'_2$, then $PP'=C_1 R_z(\alpha/2)C''R_z(\alpha'/2)C'_2$, where $C''=C_2C'_1$. We can optimize the CNOT circuit $C''$ with the CNOT circuit optimization algorithms~\cite{jiang2020optimal}. Actually, the $C''=C_2C'_1=\mathcal{L}_Q'$ where $Q'=\{j|s_j\oplus s'_j=1\}$. Combining more phase gadgets is the same as that of 2 phase gadgets. We can decompose $k$ phase gadgets into a circuit $C_1 R_z(\alpha_1/2) C_2R_z(\alpha_2/2)\cdots  C_kR_z(\alpha_k/2)$. 
    
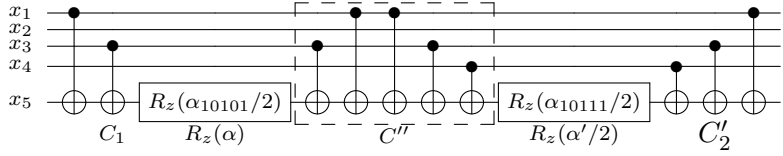
\begin{figure*}[!htbp]
    \begin{center}
    \mbox{
        \Qcircuit @C=.5em @R=0.45em {
        \lstick{\scriptstyle x_1}&\ctrl{4}&\qw&\qw&\qw&\ctrl{4}&\ctrl{4}&\qw&\qw&\qw&\qw&\qw&\ctrl{4}&\qw\\
        \lstick{\scriptstyle x_2}&\qw&\qw&\qw&\qw&\qw&\qw&\qw&\qw&\qw&\qw&\qw&\qw \gategroup{1}{5}{5}{9}{.7em}{--}&\qw\\
        \lstick{\scriptstyle x_3}&\qw&\ctrl{2}&\qw&\ctrl{2}&\qw&\qw&\ctrl{2}&\qw&\qw&\qw&\ctrl{2}&\qw&\qw\\
        \lstick{\scriptstyle x_4}&\qw&\qw&\qw&\qw&\qw&\qw&\qw&\ctrl{1}&\qw&\ctrl{1}&\qw&\qw&\qw\\
        \lstick{\scriptstyle x_5}&\targ&\targ&\gate{\scriptstyle R_z(\alpha_{10101}/2)}&\targ&\targ&\targ&\targ&\targ&\gate{\scriptstyle R_z(\alpha_{10111}/2)}&\targ&\targ&\targ&\qw\\
         & &\scriptstyle C_1&\scriptstyle R_z(\alpha)&&&\scriptstyle C''&&& \scriptstyle R_z(\alpha'/2)&&C'_2&&\\
        }}
    \end{center}%
    \caption{\centering{The combination of phases gadget.}}
    \label{fig:comb_phase}
\end{figure*}

Since phase gadgets are diagonal unitary matrices, any two phase gadgets are commute. We can use this property to reduce the quantum costs of phase gadget circuits with proper order.

The phase gadget is helpful in the qubit circuit synthesis. In 2005, Bergholm et al. proved that any $n$-qubit uniformly controlled $R_z$ gate(UCG) can be decomposed to $O(2^n)$ phase gadgets~\cite{bergholm2005quantum}. We can further synthesize more general unitary and prepare any quantum state on a qubit system using these circuits.

\section{Algorithm}
\label{sec:quditphasegadget}
We now introduce the phase gadget method in the qudit system. Further, we will focus on using the phase gadget method to reduce the circuit depth and synthesize under connectivity restrictions.



\subsection{Qudit diagonal unitary synthesis algorithm}

The diagonal unitary synthesis algorithm can be decomposed into two parts: parameters transform and gate implementation. We show the correctness of our algorithm in the Lemma \ref{lem:trans} by showing the transform between the qudit diagonal unitary and the qudit phase gadget circuit. We introduce the detailed parameters transform between the diagonal unitary and the parameters in the qudit phase gadget in the Section \ref{sec:paratrans}. Then we adjust the list of phase gadgets that any adjacent two phase gadgets differ from only one bit, which reduces the number of $SUM_d$ gates; the detailed design of the list of qudit phase gadgets is shown in Section \ref{sec:lispha}. We give a pseudo-code in the following algorithm.
\begin{algorithm}[ht]
  \SetStartEndCondition{ }{}{}%
\SetKwProg{Fn}{def}{\string:}{}
\SetKwFunction{Range}{range}
\SetKw{KwTo}{in}\SetKwFor{For}{for}{\string:}{}%
\SetKwIF{If}{ElseIf}{Else}{if}{:}{else if}{else:}{}%
\SetKwFor{While}{while}{:}{fintq}%
\AlgoDontDisplayBlockMarkers\SetAlgoNoEnd\SetAlgoNoLine%
  
  \SetKwFunction{Clause}{\bf Clause}
  \SetKwFunction{MCT}{\bf MCT}
  \SetKwProg{Fn}{}{:}{}
  \SetKwInOut{Input}{input}
  \SetKwInOut{Output}{output}
  \Input{$n$-qudit diagonal unitary $U\in \mathbb{R}^{d^n\otimes d^n}$.}
  \Output{A circuit $\Ccal$ such that $\Ccal\ket{j}=U_{jj}\ket{j}$ for any $j\in d^n$.}
  \BlankLine
  $B:= [0]$; \quad\tcp{Define the matrix $B$.}
  \For{$i$ in 0 to $d^n - 1$}{
    \For{$j$ in 0 to $d^n - 1$}{
        $t\gets \langle s_i,s_j\rangle$\;
        \If{t == 0}{
            $B_{ij}\gets -\frac{1}{d^{n-1}}$\;
        }\ElseIf{t == 1}{
            $B_{ij}\gets\frac{1}{d^{n-1}}$\;
        }\tcp{$B_{ij}=\frac{1}{d^{n-1}}(\delta_1(\langle s_i,s_j\rangle)-\delta_0(\langle s_i,s_j\rangle))$.}
    }
    $\beta_i :=U_{ii}$; \quad\tcp{Get the vector $\beta_U$.}
  } 
  $\alpha_U := B\beta_U$; \quad\tcp{Parameters transform.}
  
    $s\gets 0$\;
    \For{$i$ in 0 to $n-1$}
    {
        $s \gets s \oplus d^i$\;
        Apply gate $R_{d,z}(\alpha_{U,s})$\;
        \For{$j$ in 0 to $d^i - 1$}{
            \For{$k$ in 0 to i-1}{
                \If{$(j+1) \mod d^{k} != 0$}{
                    break\;
                }
            }
            \tcp{Find the minimal $k$ that $d^{k} \nmid j+1$}
            Apply gate $SUM_d$, where the target qubit is $q_i$ and the control qubit is $q_{k+1}$\;
            $s\gets s \oplus d^{k+1}$\;
            Apply gate $R_{d,z}(\alpha_{U,s})$; \quad\tcp{The $s$-th term of $\alpha_U$.}
        }
        $s \gets s \oplus d^i$\;
    }

\caption{$n$-qudit diagonal unitary synthesis algorithm}\label{alg:quditphasegaeget} 
\end{algorithm}

\subsection{Qudit phase gadget circuit method}
\label{sec:paratrans}
The phase gadget on the {$n$-qudit} system $P_{\alpha,s,t}$ can be defined as follows. For any $\alpha\in\mathbb{R}, s\in[d]^n,t\in[d],$ the phase gadget $P_{\alpha,s,t}$ is a circuit which can transfer the {$n$-qudit} state $\ket{x}$ to $e^{i\alpha\delta_t(<x,s>)}\ket{x}$, where the $\delta_t(y)=1$ if $y=t$ and $\delta_t(y)=0$ for other $y$. And the following equation holds: $e^{i\alpha\Pi Z_j^{s_j}}=\Pi_{j\in[d]}P_{\alpha^j,s,j}$.
Similarly, the qudit phase gadget can be decomposed into $SUM_d$ and $R_z$ gates. Here we also divide the qudit phase gadget $P_{\alpha,s,t}$ into 3 parts, i.e., 
{$S_1,R_z(\alpha),S_2$}, where $S_1,S_2$ are $SUM_d$ circuits, $S_1=\mathcal{L}_s$ and $S_2=S_1^{-1}$. The $R_z(\alpha)=\mbox{diag}(1,\cdots,1,e^{i\alpha},1,\cdots,1)$. 

\begin{figure}[!htbp]
    \begin{center}
    \mbox{
        \Qcircuit @C=.7em @R=0.7em {
        \lstick{\ket{x_1}} & \qw & \qw &\ar@{--}[]+<1.25em,1em>;[dddddddd]+<1.25em,-1em> \qw & \qw &\ar@{--}[]+<-1.25em,1em>;[dddddddd]+<-1.25em,-1em> \qw & \qw  & \qw & \qw\\
        \lstick{\vdots~~~} & \qw & \qw & \qw & \qw & \qw& \qw & \qw & \qw\\
        \lstick{\ket{x_{l_1}}} & \gate{s_{l_1}} \qwx[6] & \qw & \qw & \qw & \qw & \qw & \gate{s_{l_1}} \qwx[6]& \qw\\
        \lstick{\vdots~~~} & \qw & \qw & \qw & \qw& \qw& \qw& \qw& \qw\\
        \lstick{\ket{x_{l_2}}} & \qw& \gate{s_{l_2}} \qwx[4] & \qw & \qw & \qw& \gate{s_{l_2}} \qwx[4] & \qw& \qw\\
        \lstick{\vdots~~~} & \qw & \qw & \qw& \qw& \qw& \qw & \qw & \qw\\
        \lstick{\ket{x_{l_{|l|}}}} & \qw & \qw & \gate{s_{l_{|l|}}} \qwx[2] & \qw & \gate{s_{l_{|l|}}} \qwx[2]& \qw & \qw& \qw\\
        \lstick{\vdots~~~} & \qw &\dstick{~~~~~\cdots} \qw& \qw & \qw & \qw &\dstick{\cdots~~~~~} \qw & \qw& \qw\\
        \lstick{\ket{x_{j}}} &  \targ & \targ & \targ & \gate{ \scriptstyle R_{d,z}(\alpha)}& \gate{} &\gate{} & \gate{} & \qw\\
          && S_1 &  & { \scriptstyle R_z(\alpha)}&   &S_2 &  &  \\
        }}
    \end{center}
    \caption{{The qudit phase gadget. And the two qudit gates in $S_2$ are the inverse of the $SUM_d$ gate, the number in the square menas we repeat the $SUM_d$ or $SUM_d^{-1}$ $s_{l_i}$ times. Here $j\in l$.}}
    \label{fig:quditcnotcircuit}
\end{figure}
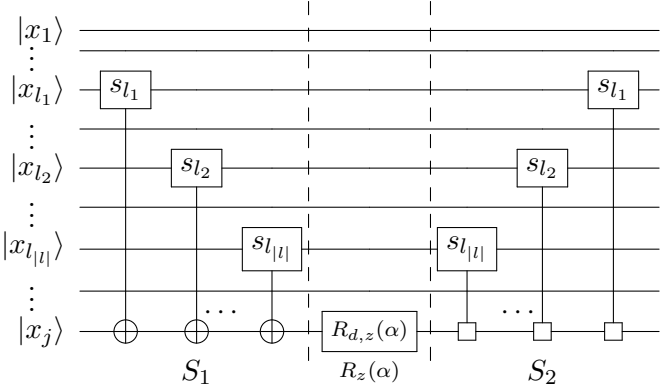

We can also combine two different phase gadgets on the qudit system. Suppose we want to implement $ P_1=P_{\alpha_1, s_1, t_1}$ and $P_2=P_{\alpha_2, s_2, t_2}$ on the qudit system. If $s_1=s_2$, then the $S_2$ in $P_1$ and the $S_1$ in $P_2$ can be eliminated. 
If $s_1\ne s_2$ but $s_1=ks_2,k\in[d]$, $P_{\alpha_2,s_2,t_2}=P_{\alpha_2,ks_2,kt_2}=P_{\alpha_2,s_1,kt_2}$, then the situation is same as the first one.
If $s_1\ne k s_2,k\in[d]$, let $s'=s_1- s_2$, the $S_2$ in $P_1$ and the $S_1$ in $P_2$ can be optimized to a linear $SUM_d$ circuit $\mathcal{L}_{s'}$. 
Some other properties hold on the qudit system, such as the phase gadgets commute. Thus, we can combine the phase gadgets with the same pair $(s,t)$ into one. 
For $n$ qudit system, any qudit phase gadget $P_{\alpha,s,t}$ can convert to : 1) a qudit phase gadget $P_{\alpha,inv(t)s,1}$ when $t\ne 0$, where $inv(\cdot)$ is the inverse in the $\mathbb{F}_d$; 2) $d-1$ phase gadgets and a global phase $e^{i\alpha}\Pi_{j=1}^{d-1}P_{-\alpha,js,1}$ when $t=0$.

Now, we will start synthesizing diagonal unitary matrices with the phase gadgets. Let $\mathcal{S}_{n}=\{(s,1)|s\in[d]^n,s\ne0^n\}$, and any phase gadget circuit can be transferred to a phase gadget circuit with at most $|\mathcal{S}|=d^n-1$ phase gadgets. We use a vector $\alpha=(\alpha_1,\alpha_2,\cdots,\alpha_{d^n-1})^T$ to record the rotation angle of these phase gadget. For a diagonal unitary $\Lambda_{d,n}=e^{i\beta_0}\mbox{diag}(1,e^{i\beta_1},e^{i\beta_2},\cdots,e^{i\beta_{d^n-1}})$, we use the vector $\beta_\Lambda=(\beta_1,\beta_2,\cdots,\beta_{d^n-1})^T$ to note the phase of the diagonal unitary. In the qubit system, we can always find an $2^n$ vector $\alpha_\Lambda$ for any $n$ qubit diagonal unitary, such that $\Pi_{j=1}^{2^n-1}P_{\alpha_{\Lambda j},s_j}=\Lambda_{2,n}$. We will prove that in the qudit system, we can decompose any diagonal unitary into at most $d^n-1$ phase gadgets.

\begin{lemma}
Any $n$ qudit diagonal unitary $\Lambda_{d,n}$ can be decomposed to at most $d^n-1$ phase gadgets, which means $\Lambda_{d,n}=\Pi_{j=1}^{d^n-1}P_{\alpha_j,s_j,t_j}$, where $s_j\in\mathcal{S}$. And the transform between $\alpha_\Lambda$ and $\beta_\Lambda$ can be calculated in $O(d^n)$ classical running time.
\label{lem:trans}
\end{lemma}
\textbf{Proof}
This proof is a constructional proof for this lemma.
We will prove this lemma by giving the transform between two vectors for any diagonal unitary.

According to the previous analyses, the phase gadgets are commuted, W.L.O.G, and we reorder the phase gadgets by lexicographic order. That is $t_j=1,s_j=(j+1)_d,j\in[d^n-1]$, where $(\cdot)_d$ means the $d$-ary representation of $(\cdot)$.
Since a phase gadget $P_{\alpha,s,t}$ can transfer the state $\ket{x}$ to $e^{i\alpha\delta_1(\langle x,s\rangle)}$, then the matrix representation of $P_{\alpha,s,t}$ is a diagonal matrix, and $d^{n-1}$ items are $e^{i\alpha}$ and the rest are $1$. This analysis allows us to construct the transform from $\alpha_\Lambda$ to $\beta_\Lambda$.
\[\beta_\Lambda=A\alpha_\Lambda,A=[A_{ij}],i,j\in[d^n-1],\]
where
\[A_{ij}=\delta_{1}(\langle s_i,s_j\rangle).\]

We now prove that $A$ is invertible. Then by $\alpha_\Lambda={A}^{-1}\beta_\Lambda$, we can find a vector $\alpha_\Lambda$ for any diagonal unitary $\Lambda_{d,n}$, such that $\Lambda_{d,n}=\Pi_{j=1}^{d^n-1}P_{\alpha_j,s_j,t_j}$ .

Define the matrix $B=[B_{ij}]$ for $i,j\in[d^n-1]$, where
\[B_{ij}=\frac{1}{d^{n-1}}\left(\delta_{1}(\langle s_i,s_j\rangle)-\delta_{0}(\langle s_i,s_j\rangle)\right).\]
Then for the matrix $C=AB$, $C_{ij}=\sum_{k=0}^{d^n-2}A_{ik}B_{kj}$. When $i=j$, 
\begin{align}
C_{ii}&=\sum_{k=0}^{d^n-2}A_{ik}B_{ki}\\
&=\sum_{k=0}^{d^n-2}\frac{1}{d^{n-1}}(\delta_{1}(\langle s_i,s_k\rangle))^2\\
&=\sum_{k=0}^{d^n-2}\frac{1}{d^{n-1}}(\delta_{1}(\langle s_i,s_k\rangle))=1.
\end{align}

When $i\ne j$, now we prove the following equation holds for any $i,j\in[d^n-1]$:
\begin{align}
C_{ij}=\sum_{k=0}^{d^n-2}A_{ik}B_{kj}=0.
\end{align}

\begin{enumerate}
    \item For the case $t s_i=s_j,t\in\{2,3,\cdots,d-1\}$, when $\delta_1(\langle s_i,s_k\rangle)=1$, then $\langle s_j,s_k\rangle=t\langle s_i,s_k\rangle=t$, that is, $\delta_{1}(\langle s_j,s_k\rangle)=\delta_{0}(\langle s_j,s_k\rangle)=0$. 
    \item For the case $t s_i\ne s_j,\forall t\in\{2,3,\cdots,d-1\}$, then we can always find two indices $u,v\in [n]$ that $s_i(u)s_j(v)\ne s_i(v)s_j(u)$, where $s_*(\cdot)$ means the $(\cdot)$-th item in the vector $s_*$. The equation $s_i(u) x+s_i(v)y=0$ has $d$ solutions, $(1,(d-s_i(u))inv(s_i(v))),$ $(2,(d-2s_i(u))inv(s_i(v)))\cdots$. Let $\mathcal{X}=\{(x,y)|s_i(u) x+s_i(v)y=0\}$, for any two different pairs $(x_1,y_1),(x_2,y_2)\in\mathcal{X}$, the values of $s_j(u) x+s_j(v)y$ takes and only takes all the values from 0 to $d-1$ once. Let $P_s=\{t|\langle s,t\rangle=1\}$, notice that when we fixed the value of the $t$ except the place $u,v$. There are $d$ items in $P_{s_i}$, and the inner product values among these items in $s_j$ take and only take all the values from 0 to $d-1$ once. Thus
    \begin{align}
        &\sum_{k=0}^{d^n-2}\delta_{1}(\langle s_i,s_k\rangle)\delta_{1}(\langle s_j,s_k\rangle)\\
        = &\sum_{k=0}^{d^n-2}\delta_{1}(\langle s_i,s_k\rangle)\delta_{0}(\langle s_j,s_k\rangle)\\=&d^{n-2}.
    \end{align}
    
\end{enumerate}

\subsection{Construction of qudit phase gadget circuit}
\label{sec:lispha}
As we analyzed in the previous subsection, any phase gadget circuit can be decomposed to $S_1R_{d,z}(\alpha_1)S_2R_{d,z}(\alpha_2)\cdots$. In this subsection, we focus on synthesizing general diagonal unitary matrices. Thus, W.L.O.G, we assume the phase gadget circuit contains all the pairs in $\mathcal{S}_n$.

\begin{figure*}[!htbp]
    \begin{center}
    \mbox{
        \Qcircuit @C=.55em @R=0.75em {        
        \lstick{\ket{x_1}} & \qw & \qw &\ar@{--}[]+<1.25em,1em>;[dddddddd]+<1.25em,-1em> \qw &\qw & \qw &\ar@{--}[]+<-1.25em,1em>;[dddddddd]+<-1.25em,-1em> \qw & \qw  & \qw & \qw\\
        \lstick{\vdots~~~} & \qw & \qw & \qw & \qw & \qw&\qw & \qw & \qw & \qw\\
        \lstick{\ket{x_{l_1}}} & \gate{s_{l_1}} \qwx[6] & \qw &\qw & \qw & \qw & \qw & \qw & \gate{s_{l_1}} \qwx[6]& \qw\\
        \lstick{\vdots~~~} & \qw & \qw & \qw & \qw &\qw& \qw& \qw& \qw& \qw\\
        \lstick{\ket{x_{l_2}}} & \qw& \gate{s_{l_2}} \qwx[4] & \qw &\qw & \qw & \qw& \gate{s_{l_2}} \qwx[4] & \qw& \qw\\
        \lstick{\vdots~~~} & \qw & \qw & \qw&\qw & \qw& \qw& \qw & \qw & \qw\\
        \lstick{\ket{x_{l_{|l|}}}} & \qw & \qw & \gate{s_{l_{|l|}}} \qwx[2] &\qw & \qw & \gate{s_{l_{|l|}}} \qwx[2]& \qw & \qw& \qw\\
        \lstick{\vdots~~~} &\qw &\dstick{~~~~~\cdots} \qw&\qw & \qw & \qw & \qw &\dstick{\cdots~~~~~} \qw & \qw& \qw\\
        \lstick{\ket{x_{j}}} &  \targ & \targ & \targ & \gate{ \scriptstyle R_{d,z}(\alpha_1)}& \gate{ \scriptstyle R_{d,z}(\alpha_2)}& \gate{} &\gate{} & \gate{} & \qw\\
          && S_1 &  & { \scriptstyle R_z(\alpha_1)}& { \scriptstyle R_z(\alpha_2)}&   &S_2 &  &  \\
        }
        }
    \end{center}
    \caption{{Combination of two phase gadgets where $s_1=ks_2$. Notice that the $R_{d,z}$ gate can further combine into one. And the two qudit gates in $S_2$ are the inverse of the $SUM_d$ gate, the number in the square menas we repeat the $SUM_d$ or $SUM_d^{-1}$ $s_{l_i}$ times. Here $j\in l$.}}
    \label{fig:qudittwophase}
\end{figure*}
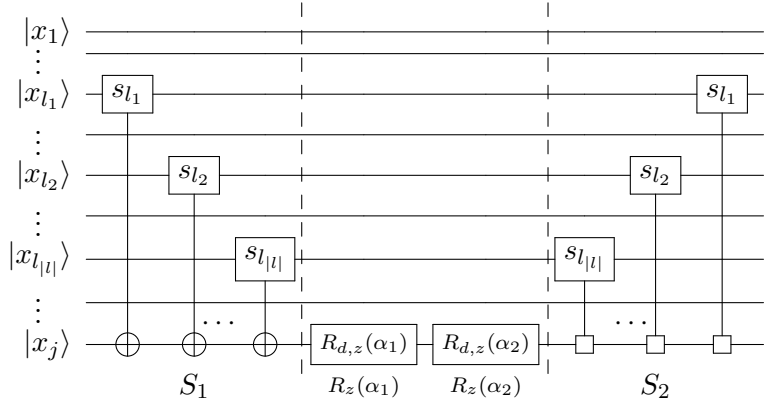

\begin{lemma}
Any qudit phase gadget circuit can be decomposed to a quantum circuit with at most $(d^{n}-1)/(d-1)$ $SUM_d$ gates and $(d^{n}-1)/(d-1)$ $R_{d,z}(\cdot)$ gates. 
\end{lemma}
\textbf{Proof.}
    In the section \ref{sec:paratrans}, we have proved that any phase gadget circuit can be transferred to a new phase gadget circuit, which contains at most $d^n-1$ phase gadgets. The $SUM_d$ circuit between two phase gadgets can be optimized to a linear $SUM_d$ circuit. Moreover, for the phase gadgets with string $s_1,s_2$ such that $s_1=ks_2,k\in[d]$, the $SUM_d$ circuit between two phase gadgets can be removed. We give an example in Fig.\ref{fig:qudittwophase}, these two $R_{d,z}$ gates can further combine into a single-qudit diagonal gate $\Lambda_{d,1}=\mbox{diag}(e^{i\alpha_0},e^{i\alpha_1},\cdots, e^{i\alpha_{d-1}})$. Thus, for any $s_1\in\mathbb{F}_d^n$, the phase gadget with string $s_2=ks_1,k\in[d]$ can be absorbed by the phase gadget with string $s_1$.

Let $\Acal_{n, i}=\{(s10^{i},t)|s\in\mathbb{F}_{d}^{n-1-i},t\in[d],t\ne 0\}$, where $\Acal_{n, i}$ denotes the phase gadget pair whose last non-zero item is 1 at place $n-i-1$. Let $\mathcal{S}_{n, i}=\{(sx0^{i},1)|s\in\mathbb{F}_{d}^{n-1-i},x\in [d],x\ne 0\}$, where $\mathcal{S}_{n, i}$ denotes the phase gadget pair whose last non-zero item is at place $n-i-1$ and $t=1$. In the section \ref{sec:paratrans}, we have proved that the phase gadget of pair $(s_a,t)$ can be replaced by a phase gadget with pair $(ks_a, kt)$, where $k\in[d]$ and $k\ne 0$. For any pair $(s,t)\in\Acal_{n,i}$, $(inv(t)s,1)$ is in the set $\mathcal{S}_{n,i}$. And for any pair $(sx0^i,1)\in \mathcal{S}_{n,i}$, $(inv(x)(sx0^i),inv(x))\in \mathcal{A}_{n,i}$. Thus, the phase gadget circuit with pair set $\Acal_{n, i}$ and the phase gadget circuit with pair set $\mathcal{S}_{n, i}$ is ``equivalent''.\footnote{Here equivalent means that we can easily transfer one phase gadget with pair set $\Acal$ to a phase gadget with pair set $\Bcal$ without influence the function of circuit.} Notice, $\lvert \mathcal{A}_{n,i} \rvert=\mathcal{S}_{n,i}=(d-1)\cdot d^{n-i-1}$. In the subsequent proof, we construct the circuit $A_{n,i}$ which contains all the phase gadgets with pair in $\Acal_{n,i}$, then any diagonal unitary can be implemented by the circuit $A_{n,1}A_{n,2}\cdots A_{n,n-1}$.

Now we give a synthesis algorithm for $A_{n, i}$. According to the properties of qudit phase gadget analyzed in the section \ref{sec:paratrans}, $A_{n,i}=S_1(\Pi_{j=0}^{\lvert \mathcal{A}_{n, i} \rvert-1}R_{d,z}(\alpha_j)S'_{j})S_2$. As shown in Fig.\ref{fig:qudittwophase}, when the string $s$ of the phase gadget is the same, the $S'_j$ can be reduced completely. We combine the phase gadgets with the same string $s$, then $A_{n,i}=S_1(\Pi_{k=0}^{d^{n-i-1}-1}\Lambda_{d,1}S'_{k})S_2$, where the parameters of $\Lambda_{d,1}$ can be easily determined by the absorbed phase gadgets. 
In \cite{bergholm2005quantum,sun2021asymptotically}, they show that the Gray Code can be used to optimize the order of phase gadgets. Using the same idea, we can use a $d$-ary Gary Code to reduce the $SUM_d$ gate number of each $S'_k$ and $S_1, S_2$ to $1,0,1$, respectively. Thus, for any $A_{n,i}$, it costs at most $d^{n-1-i}$ $SUM_d$ gates and  $d^{n-1-i}$ $R_{z,d}$ gates. Then the total cost of diagonal unitary matrix is $\sum_{i=0}^{n-1}d^{n-1-i}=(d^n-1)/(d-1)$ $SUM_d$ gates and  $(d^n-1)/(d-1)$ $R_{z,d}$ gates.

\begin{lemma}
\label{lem:dnGraycode}
    There exits a $d$-nary and $n$-digit Gary Code that the difference between two successive values is at most 1.
\end{lemma}
\textbf{Proof}
    Define the code $C(b)$ be the $d$-nary and $n$-digit Gary code. The code is constructed by the following method. $C(b)\oplus C(b+1)=e_{k}$, here $e_k=00\cdots 010\cdots00$ is a vector with $k$-th is 1 and the other is 0 and the $k$ is the minimal number such that $3^k \mid b$ and $3^k \nmid b$. For example, (00, 01, 02, 12, 10, 11, 21, 22, 20, 00) is a 3-nary and 2-digit.

\subsection{Low-depth circuit synthesis for diagonal unitary}
\label{sec:lowdepth}
In the \cite{sun2021asymptotically}, a parallel Gray code can parallel the phase gadget circuit. Their work fixed the target qubits in the phase gadget circuit (half of the ancillary qubits). The phase gadgets applied to each target qubit are listed in Gray code. And other qubits are used to reduce the cost between two phase gadget circuits.

For $d$-level systems, we define a similar but not equal code, parallel d-level Gray code. Define the code $C(a,b)$ as the $b$-th phase gadget string at the $a$-th target qudit. $C(0,b)$ is defined as in the proof of lemma \ref{lem:dnGraycode}. And then $C(a,b)\oplus C(a,b+1) \equiv (a+ C(0,b)\oplus C(0,b+1))\mod d$. The circuit depth of diagonal unitary can be reduced to $O(\frac{d^{n-1}}{n+m}+n\log d)$.

\subsection{The numerical result in section \ref{sec:lowdepth}}

We randomly sampled several $n$-qutrit diagonal unitary matrices to evaluate our algorithm described in the section \ref{sec:lowdepth}. The following steps generate the random diagonal unitary. Independently and uniformly select $3^n -1 $ random numbers $\alpha_1,\alpha_2,\cdots,\alpha_{3^n-1}$ from $[0,2\pi]$. Then $\mbox{diag}(1,e^{i\alpha_1},e^{i\alpha_2},\cdots,e^{i\alpha_{3^n-1}})$ is a random diagonal unitary.

\begin{figure}
    \centering
    \resizebox{0.45\textwidth}{!}
    {%
    \includegraphics{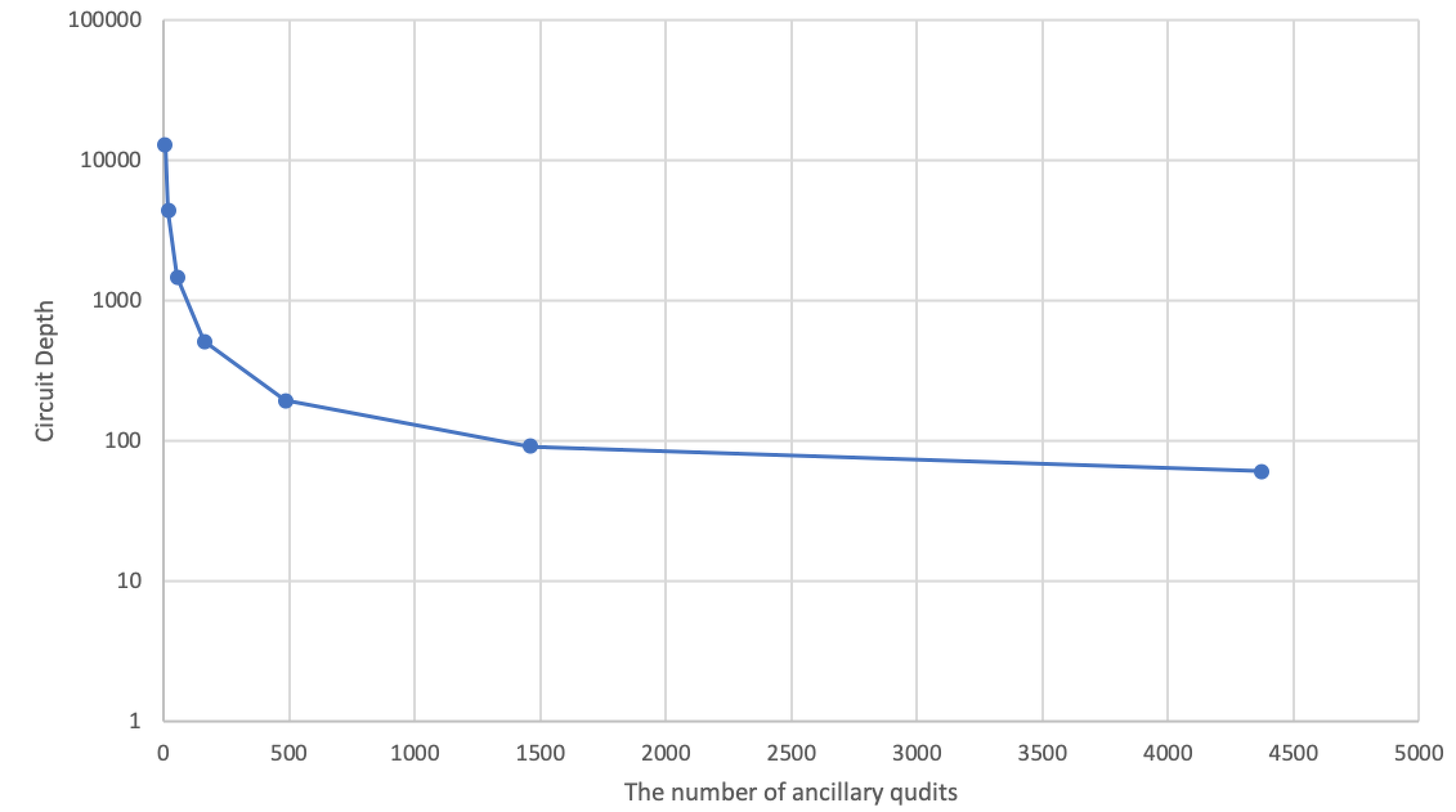}
    }
    \caption{The circuit depth with different number of ancillary qutrit for $10$-qutrit diagonal unitary.}
    
    \label{fig:anci_depth}
\end{figure}

We first focus on the $10$-qutrit diagonal unitary and randomly generate 20 samples to evaluate the performance of our algorithm with different ancillary qutrits. Fig.\ref{fig:anci_depth} shows the circuit depth with different ancillary qutrits. The circuit depth reduces rapidly with the help of ancillary qubits. This result indicates that there is a trade off between the time-space resource in the quantum computation. For $10$-qutrit diagonal unitary, our algorithm can execute in several seconds on a laptop.

\begin{figure}
    \centering
    \resizebox{0.45\textwidth}{!}{%
    \includegraphics{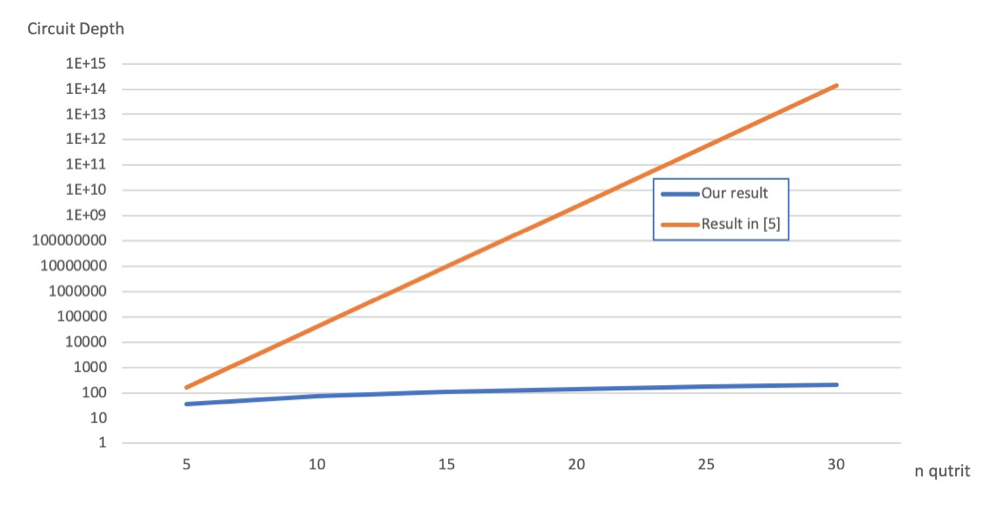}
    }
    \caption{The circuit depth comparation between our algorithm and algorithm in \cite{Bullock2004AsymptoticallyOQ}}
    \label{fig:log_depth}
\end{figure}

We then assume enough ancillary qutrits is allowed (about $2\times d^n$ qutrits) and compare our result with the circuit depth in \cite{Bullock2004AsymptoticallyOQ}. We fix the level of the system $d=3$ and choose the number of qutrits ranging from $5$ to $25$. The result when $n=25$ takes several days on a laptop, and other results can be quickly obtained in several seconds on a laptop. The results are shown in Fig. \ref{fig:log_depth}. The circuit depth of our algorithm grows slower than the algorithm in \cite{Bullock2004AsymptoticallyOQ}. Our algorithm makes it possible to reduce the circuit depth to a low level with ancillary qudits.

\subsection{Diagonal unitary synthesis under connectivity restriction}
\label{sec:dia_uni_connect}
\begin{figure}
    \centering
    \includegraphics{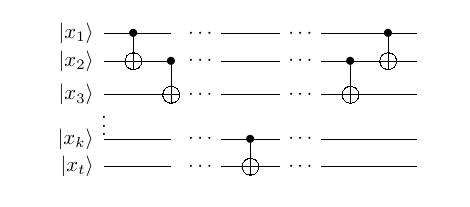}
    \caption{A stair-like circuit.}
    \label{fig:Stcicruit}
\end{figure}

In \cite{yang2024quantum}, they found that by replacing the long-distance CNOT gate with a stair-like CNOT circuit, as shown in Fig.\ref{fig:Stcicruit}, the synthesis algorithm can be used for synthesis under connectivity restriction.

In this paper, we use the ``center finding'' method, choose a ``center'' qudit as the target qudit, and reorder other qudits by the distance to the center qudit. Then, the number of long-distance $SUM$ gates can be reduced rapidly. Finally, the stair-like circuit replaces the long-distance $SUM$ gate. For a $d$-level system, the left (including the middle one) of the stair-like circuit is $SUM$ gate, and the right of the stair-like circuit is $SUM^{-1}$ gate. The circuit size of the diagonal unitary synthesis under connectivity restriction is still $O(d^{n-1})$, where the mapping method may cost $O(nd^{n-1})$.

\section{Application of qudit phase gadget}
\label{sec:application}
\subsection{Qudit state preparation}
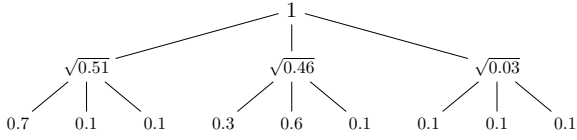
\begin{figure}[!ht]
\centering
    \centering
    \resizebox{0.45\textwidth}{!}{%
    \begin{tikzpicture} [level distance=1cm,
        level 1/.style={sibling distance=3.6cm},
        level 2/.style={sibling distance=1.2cm},]
        \node {$1$}
            child {node[scale=0.8] {$\sqrt{0.51}$}
                child {node[scale=0.8] {$0.7$} }
                child {node[scale=0.8] {$0.1$}}
                child {node[scale=0.8] {$0.1$}}
            }
            child {node[scale=0.8] {$\sqrt{0.46}$}
                child {node[scale=0.8] {$0.3$} }
                child {node[scale=0.8] {$0.6$}}
                child {node[scale=0.8] {$0.1$}}
            }
            child {node[scale=0.8] {$\sqrt{0.03}$}
                child {node[scale=0.8] {$0.1$} }
                child {node[scale=0.8] {$0.1$}}
                child {node[scale=0.8] {$0.1$}}
            };
    \end{tikzpicture}%
    }
    \label{fig:bst}
\label{fig:circuit_psi} 
\caption{ Tree representation for $\ket{\psi}=(0.7,0.1,0.1,0.3,0.6,0.1,0.1,0.1,0.1)^T$. 
}
\label{fig:2-qutrit}
\end{figure}

The state preparation problem is a fundamental problem in quantum compilation since state preparation is always the first step of some quantum computation algorithms. 
In this section, we adopt a method that is similar but not completely identical to the two-level system in order to prepare any $n$-qudit quantum state. The state preparation algorithm is based on the $d$-ary tree. We can use a $n$-layer $d$-ary tree to represent a $n$ qudit state, where the leaf nodes store the amplitudes of the state, and the inner nodes store the square root of the sum of squared amplitudes of all subtrees. We give an example of a $2$ qutrit state and its tree representation. Let $\ket{\psi}=(0.7,0.1,0.1,0.3,0.6,0.1,0.1,0.1,0.1)$, then the corresponding tree is shown in Fig.~\ref{fig:2-qutrit}.

We now introduce a qudit-by-qudit state preparation framework for $n$-qudit state. The pseudocode of the qudit state preparation algorithm is shown in Alg.~\ref{alg:quditqsp}.
\FloatBarrier

\begin{algorithm}[t]
  \SetStartEndCondition{ }{}{}%
\SetKwProg{Fn}{def}{\string:}{}
\SetKwFunction{Range}{range}
\SetKw{KwTo}{in}\SetKwFor{For}{for}{\string:}{}%
\SetKwIF{If}{ElseIf}{Else}{if}{:}{else if}{else:}{}%
\SetKwFor{While}{while}{:}{fintq}%
\AlgoDontDisplayBlockMarkers\SetAlgoNoEnd\SetAlgoNoLine%
  
  \SetKwFunction{Clause}{\bf Clause}
  \SetKwFunction{MCT}{\bf MCT}
  \SetKwProg{Fn}{}{:}{}
  \SetKwInOut{Input}{input}
  \SetKwInOut{Output}{output}
  \Input{$n$-qudit vector $\ket{\psi}=(\psi_0,\psi_1\cdots)\in \mathbb{C}^{d^n}$.}
  \Output{A circuit $\Ccal$ such that $\Ccal\ket{0}=\sum_{j}\psi_j\ket{j}$.}
  \BlankLine
  \tcp{A qudit-by-qudit algorithm, In this algorithm we prepare the state according to the tree representation. The algorithm can be divided into $n$ steps. In step $\ell$, we distribute the amplitude of the $\ell$-layer node into the $\ell+1$-layer node.}
  \For{$\ell$ in $0$ to $n-1$}{
    \tcp{In each step, we implement $\lceil \log{d}\rceil$ diagonal unitary and $O(\log d)$ single qudit gates.} 
      \For{$j$ in $0$ to $\lceil \log{d}\rceil-1$}{
        \tcp{Distribute the amplitude, that is $\sqrt{\psi_k^2+\psi_{k+2^j}^2}\ket{k}\to {\psi_k}\ket{k}+\psi_{k+2^j}\ket{k+2^j}$, for all $k\in[2^j]$.}
        Apply $R_{d,z}(\pi/2^{j+1},2\pi/2^{j+1},\cdots,(d-1)\pi/2^{j+1})$ on qudit $q_\ell$\;
        Apply all $H_{d,k,k\oplus 2^{j}}$ for all $k\in[2^{j}]$ on qudit $q_\ell$\;
        Apply a diagonal unitary $U$ on qubit $\ell$\;
        \tcp{The $U$ can be calculated in Alg.\ref{alg:quditU}.}
        Apply all $H_{d,k,k\oplus 2^{j}}$ for all $k\in[2^{j}]$ on qudit $q_\ell$\;
        Apply $R_{d,z}(\pi/2^{j+1},2\pi/2^{j+1},\cdots,(d-1)\pi/2^{j+1})$ on qudit $q_\ell$\;
      }

  }

\caption{$n$-qudit state preparation algorithm framework}\label{alg:quditqsp} 
\end{algorithm}
For the example in Fig.~\ref{fig:2-qutrit}, the algorithms contains two steps. The first step makes $\ket{0}$ to \[\sqrt{0.51}\ket{0}+\sqrt{0.46}\ket{1}+\sqrt{0.03}\ket{2}.\]
Then after the second step the state is changed to $0.7\ket{00}+0.3\ket{10}+0.6\ket{11}+0.1(\ket{01}+\ket{02}+\ket{12}+\ket{20}+\ket{21}+\ket{22}).$

Each step can be decomposed to $2$ diagonal unitary matrices and $4$ $H_d$ gates. In the first step, the first diagonal unitary matrix is $U_1=\mbox{diag}(e^{-i\theta_1 },e^{i\theta_1},1)$, where $\theta_1=\arccos{(\frac{\sum_{x=0,1,2}\psi_{0x}^2 +\sum_{x=0,1,2}\psi_{2x}^2}{\sum_{x=0,1,2}\psi_{0x}^2+\psi_{1x}^2+\psi_{2x}^2})}.$  The second diagonal unitary matrix is $U_2=\mbox{diag}(e^{-i\theta_2 },1,e^{i\theta_2})$, where $\theta_2=\arccos{(\frac{\sum_{x=0,1,2}\psi_{0x}^2}{\sum_{x=0,1,2}\psi_{0x}^2+\psi_{2x}^2})}.$ Similarly, we can calculate the diagonal unitary for the second step. Here, $U_3 = \mbox{diag}(e^{-i\theta_3},e^{i\theta_3},1,e^{-i\theta_4},e^{i\theta_4},1,e^{-i\theta_5},e^{i\theta_5},1),$
where $\theta_3=\arccos{\frac{\sqrt{0.5}}{\sqrt{0.51}}}$, $\theta_4=\arccos{\frac{\sqrt{0.45}}{\sqrt{0.46}}}$, $\theta_5=\arccos{\frac{\sqrt{0.02}}{\sqrt{0.03}}}$. And 
$U_4 = \mbox{diag}(e^{-i\theta_6},e^{i\theta_6},1,e^{-i\theta_7},e^{i\theta_7},1,e^{-i\theta_8},e^{i\theta_8},1),$
where
$\theta_6=\arccos{\frac{0.7}{\sqrt{0.5}}},\theta_7=\arccos{\frac{0.3}{\sqrt{0.45}}},\theta_8=\arccos{\frac{0.1}{\sqrt{0.02}}}.$

For a general $U$, the parameter can be computed by the Alg.~\ref{alg:quditU}.
As analyzed earlier in previous section, the circuit depth of diagonal unitary can be reduced to $O(d^{n-1}/n+m +n \log d)$.

\FloatBarrier

\begin{algorithm*}[htbp!]
  \SetStartEndCondition{ }{}{}%
\SetKwProg{Fn}{def}{\string:}{}
\SetKwFunction{Range}{range}
\SetKw{KwTo}{in}\SetKwFor{For}{for}{\string:}{}%
\SetKwIF{If}{ElseIf}{Else}{if}{:}{else if}{else:}{}%
\SetKwFor{While}{while}{:}{fintq}%
\AlgoDontDisplayBlockMarkers\SetAlgoNoEnd\SetAlgoNoLine%
  
  \SetKwFunction{Clause}{\bf Clause}
  \SetKwFunction{MCT}{\bf MCT}
  \SetKwProg{Fn}{}{:}{}
  \SetKwInOut{Input}{input}
  \SetKwInOut{Output}{output}
  \Input{$n$-qudit vector $\ket{\psi}=(\psi_0,\psi_1\cdots)\in \mathbb{C}^{d^n}$. The steps number $\ell$ and the parts number $j$. The value of the $\ell+1$-layer node $\nu=(\nu_0,\nu_1\cdots,\nu_{d^{\ell+1}-1})\in\mathbb{C}^{d^{\ell+1}}$.}
  \Output{The corresponding diagonal unitary $U$}
  \BlankLine

  $p\gets \min(d-2^j,2^j)$\;
  \For{$w$ in 0 to $d^{\ell}-1$}{
  \For{$r$ in $0$ to $p-1$}{
    $s,t\gets 0$\;
     \For{$k:=r;k<d;j=j+2^{j+1}$}{
        $s \gets s + \nu_{k+dw}^2$\;
     }
     \For{$k:=r+2^{j};j<d;j=j+2^{j+1}$}{
        $t \gets t + \nu_{k+dw}^2$\;
     }
     $\theta_{j+1}= \arccos(\sqrt{\frac{s}{s+t}})$\;
  }
  \If{$d\ge 2p$}{
    $U_w\gets \exp\left(\mbox{diag}(i\theta_1,i\theta_2,\cdots,i\theta_p,-i\theta_1,-i\theta_2,\cdots,-i\theta_p,0,0,\cdots,0)\right)$
  }\Else{
    $U_w\gets \exp\left(\mbox{diag}(i\theta_1,i\theta_2,\cdots,i\theta_p,0,0,\cdots,0,-i\theta_1,-i\theta_2,\cdots,-i\theta_p)\right)$
  }
  }
  $U\gets[U_0,U_1,\cdots,U_{d^{\ell}-1}]$
     
\caption{Diagonal unitary $U$ in the Alg.~\ref{alg:quditqsp}}\label{alg:quditU} 
\end{algorithm*}


Due to the reduction of circuit depth of diagonal unitary, we now can reduce the circuit depth of qudit quantum state preparation to $O(d^{n-1}\log {d}/(n+m))$. 
\begin{corollary}
Using $SUM_d$ gates and single qubit gates, any $n$-qudit quantum state can be prepared by an $O(\frac{d^{n-1}\log d}{n+m}+n^2\log d)$-depth quantum circuit with $m$ ancillary qudits.
\end{corollary}

We can use the same framework in \cite{sun2021asymptotically} and~\cite{yuan2023optimal} to reduce the depth and size of $d-base$ qudit state preparation problem. All operations $modulo-2$ in previous quantum circuits are now $modulo-d$. For example, the function of the CNOT gate in the above two papers is like a $SUM_d$ gate in this paper.

Our method can be divided into three parts according to the different numbers of ancillary qudits. If we have enough $m$ ancillary qudits to solve Quantum State Preparation (QSP) problem, the depth of the circuit is $O\left(n(n+1)-t(t+1)+\frac{d^n-1}{n+m}+n\right)$, and the size is $O\left(d^n\right)$, where $t=\left\lfloor\log \left(\frac{\mathrm{m}}{\mathrm{d}+1}\right)\right\rfloor$. If QSP problem is transformed into Controlled Quantum State Preparation(CQSP) problem, where the number of control qubits is $k$ and for ancillary qudits is $m$, the depth of the circuit is $O(n+k+\frac{d^{n+k-1}}{n+k+m})$, and the size is $O(d^{n+k-1})$. When $k=0$, CQSP problem reduces to QSP, with the circuit depth of $\Theta(n+\frac{d^{n-1}}{n+m})$ and the size of $\Theta(d^{n-1})$. If there are no ancillary qudits in the last setting, the depth of the circuit is $O(\frac{d^{n-1}}{n})$, and the size is $O(d^{n-1})$.

The main techniques used in the above statement are as follows. We utilize unitary matrix decomposition to assess the depth and size. For QSP problem with enough ancillary qudits, we first employ unary coded state preparation and then convert it to the $d$-base coding state. The remaining part involves using diagonal gates with ancillary qudits. For the second situation, we employ control qubits and ancillary qudits to transform QSP into CQSP, which refers to Rosenthal’s quantum state preparation framework in~\cite{rosenthal2021query}. In the last setting, without ancillary qudits, we use another string arrangement scheme combined with diagonal gates to solve QSP problem recursively.

By using the counting method and the light cone method, we could find that almost all $n$-qudit quantum states can not be prepared by any $o(d^{n-1}/(m+n))$-depth quantum circuits and any $o(n\log d)$-depth quantum circuits.

\begin{corollary}
For almost all the $n$-qudit quantum states, they can not be prepared by any $o(d^{n-1}/(m+n))$-depth quantum circuits with $m$ ancillary qudits or any $o(n\log d)$-depth quantum circuits.
\end{corollary}

Our algorithm is asymptotically optimal in circuit depth. To the best of our knowledge, we first reduce the circuit depth of the qudit quantum state preparation problem to poly$(n,d)$.

This method also works when the connectivity is restricted: we can replace the algorithm \ref{alg:quditU} with the algorithm shown in section \ref{sec:dia_uni_connect} to conquer this restriction.


\subsection{Qudit general unitary synthesis and quantum Householder reflections}
The diagonal unitary is also widely used in the general unitary synthesis algorithm. A well-known algorithm, quantum Householder reflection (QHR)~\cite{Householder1958,Ivanov2006Engineering,Bullock2004AsymptoticallyOQ}, decomposed the general unitary into several quantum state preparation unitary and diagonal unitary.

We define a general QHR as
\[M(v;\varphi)=I+(e^{i\varphi}-1)|v\rangle\langle v|, v\in\mathbb{C}^{d^n},\varphi\in\mathbb{R}. \]
\begin{lemma}[\cite{Bullock2004AsymptoticallyOQ} Section 3.B]
    Any $N\times N$ unitary $U$ can be decomposed to $M(v_1;\varphi_1)M(v_2;\varphi_2)\cdots M(v_N;\varphi_N)$.
\end{lemma}

Notice that $M(v;\varphi)=I+(e^{i\varphi}-1)|v\rangle\langle v|=S_v(I+(e^{i\varphi}-1)|0\rangle\langle 0|)S_v^{\dagger}$, where $S_v$ is the state preparation unitary for $v$ that $S_v\ket{0}=\ket{v}$.
Thus, any $n$-qudit unitary can be decomposed to $2d^n$ qudit quantum state preparation circuits and $d^n$ diagonal unitary matrices.
\begin{lemma}
    Using $SUM_d$ gates and single qubit gates, any $n$-qudit unitary can be implemented by an $O(\frac{d^{2n-1}\log{d}}{n+m}+n^2d^n\log{d})$-depth quantum circuit with $m$ ancillary qudits.
\end{lemma}
The result is nearly asymptotically optimal when the number of ancillary qubits is $o(d^n)$.

When the connectivity architecture is fixed, the method shown in section \ref{sec:dia_uni_connect} can help the algorithm perform well under connectivity restrictions. For any $n$-qudit unitary $U$, our algorithms take at most $O(d^{n-1})$ elementary gates to synthesize unitary $U$, which mitigates the impact of connectivity restriction.

\section{Conclusion}
\label{sec:conclusion}
In the work, we propose a novel qudit phase gadget method. With the qudit phase gadget method, we can quickly reduce the quantum depth of diagonal unitary. We then use the qudit phase gadget method to solve the quantum qudit state preparation problem and general qudit unitary synthesis problem. Our method significantly reduces circuit depth. Combined with the lower bound result, our method can solve these synthesis problems with (almost) asymptotically optimal circuit depth. Our method serves as a bridge between diagonal unitary and phase gadgets, connecting them. This connection enables the opportunity to reduce the circuit depth on qudit system devices. Additionally, due to the properties of the phase gadget, our method may be further generalized to include Pauli gadgets, which are widely utilized in quantum chemistry simulation and quantum machine learning.
\bibliographystyle{plainnat}
\bibliography{mybibliography}
\end{document}